\documentclass[12pt]{article}
\usepackage{graphicx}

\newcommand\pubnumber{SI-HEP-2011-01}
\newcommand\pubdate{January 11, 2011}

\def\Title#1{\begin{center} {\Large #1 } \end{center}}
\def\Author#1{\begin{center}{ \sc #1} \end{center}}
\def\Address#1{\begin{center}{ \it #1} \end{center}}

\newcommand\pubblock{\rightline{\begin{tabular}{l} \pubnumber\\
         \pubdate  \end{tabular}}}
\newenvironment{Abstract}{\begin{quotation}  }{\end{quotation}}
\newenvironment{Presented}{\begin{quotation} \begin{center} 
             PRESENTED AT\end{center}\bigskip 
      \begin{center}\begin{large}}{\end{large}\end{center} \end{quotation}}
\def\Acknowledgements{\bigskip  \bigskip \begin{center} \begin{large}
             \bf ACKNOWLEDGEMENTS \end{large}\end{center}}

\def\beq{\begin{equation}}
\def\eeq#1{\label{#1}\end{equation}}
\def\eeqn{\end{equation}}

\def\beqa{\begin{eqnarray}}
\def\eeqa#1{\label{#1}\end{eqnarray}}
\def\eeqan{\end{eqnarray}}

\let\bar=\overbar

\def\Dslash{\not{\hbox{\kern-4pt $D$}}}
\def\dslash{\not{\hbox{\kern-2pt $\del$}}}

\def\msb{{\bar{\ssstyle M \kern -1pt S}}}

\begin{document}
\begin{titlepage}
\pubblock

\vfill
\Title{Form Factors and Long-Distance Effects \\in  
$B\to V(P) \ell^+\ell^-$ and $B\to V \gamma$ }
\vfill
\Author{Alexander Khodjamirian}
\Address{Theoretische Physik 1, Fachbereich Physik, Universit\"at Siegen, \\
D-57068 Siegen, Germany}
\vfill
\begin{Abstract}
I overview the hadronic input for the exclusive 
flavour-changing neutral-current $B$-decays with a vector ($V=K^*,\rho$)
or pseudoscalar ($P=K,\pi$) meson in the final state.
After presenting the current status of $B\to P,V$ form factors, 
I discuss
the estimate of  the charm-loop effect in $B\to K^{(*)} \ell^+\ell^-$ and 
$B\to K^* \gamma$.  
\end{Abstract}
\vfill
\begin{Presented}
CKM2010, the 6th International Workshop\\ on the CKM Unitarity Triangle, \\
University of Warwick, UK, 6-10 September 2010
\end{Presented}
\vfill
\end{titlepage}
\def\thefootnote{\fnsymbol{footnote}}
\setcounter{footnote}{0}

\section{Introduction}
The exclusive $B\to V(P)\,\ell^+\ell^-$ and $B\to V\gamma$ decays, with a vector ($V=K^*,\rho,...$) or pseudoscalar ($P=K,\pi,...$) meson in the final state are important for the search for flavour-changing new physics. I will 
overview the current status of the hadronic input for these decays.  
In  Standard Model (SM), the CKM 
favoured $B\to K^{(*)}\ell^+\ell^-$ decay amplitudes reduce to the matrix elements   
\begin{equation} 
 A(B \to K^{(*)} \ell^+\ell^-)=
\langle K^{(*)} \ell^+\ell^-\mid H_{eff}\mid B\rangle  
\label{eq:ampl}
 \end{equation}
of the effective Hamiltonian  
\begin{equation}
H_{eff}= -\frac{4G_{F}}{\sqrt{2}}V_{tb}V_{ts}^{*}
{\sum\limits_{i=1}^{10}} C_{i}({\mu}) O_{i}({\mu})\,,
\label{eq:Heff}
\end{equation}
where the  small $\sim V_{ub}V_{us}^*$ part is  neglected, and 
the dominant  $b\to s$ operators are  $O_{9,10}$ and $O_7$, with the Wilson coefficients
$C_9(m_b)\simeq 4.2$, $C_{10}(m_b)\simeq-4.4 $ 
and $C_7(m_b)\simeq -0.3$, respectively (for a review see, e.g., \cite{BBL}).   

The hadronic matrix elements of these operators factorize, 
e.g., the contribution of the operator $O_{9}=(\bar{s}_L\gamma_\rho b_L)(\bar{\ell}\gamma^\rho\ell)$  
to the decay amplitude  (\ref{eq:ampl})
reduces to  the hadronic matrix elements $\langle K^{(*)}(p)|  \bar{s}_L\gamma_\rho b_L|B(p+q)\rangle$, 
parameterized in terms of the $B\to K^{(*)}$ form factors depending on  $q^2$, the momentum transfer to the 
lepton pair. For the $B\to V\gamma$ decay one correspondingly needs the form factors at $q^2=0 $. 

In order to access the flavour-changing neutral current 
(FCNC) interaction 
encoded in (\ref{eq:Heff}) and to trace and/or constrain new physics, 
one has to compare the measured
exclusive decay observables with the SM predictions. 
For the latter, an accurate knowledge of the $B\to P,V$ form factors
is necessary but not sufficient. Important are also 
the contributions to the $B\to V(P) \ell^+\ell^-$ and $B\to V\gamma$ decay amplitudes due to the  operators 
$O_{1,2,...,6,8g}$, which have to be analyzed one by one
(see e.g. \cite{BFS})
and added to the dominant FCNC contributions. 
Especially important are the current-current operators 
$O^{(c)}_{1}= (\bar{s}_L\gamma_\rho c_L) (\bar{c}_L\gamma^\rho b_L)$
and  $O^{(c)}_{2}= (\bar{s}^j_L\gamma_\rho c^i_L) (\bar{c}^i_L\gamma^\rho b^j_L)$
with large Wilson coefficients $C_1(m_b)\simeq 1.1$, $C_2\simeq -0.25 $.
Combined with the e.m. interactions of quarks and leptons, they
generate $b\to s $ transitions with intermediate $c$-quark loops. The 
resulting hadronic matrix elements 
contain nonfactorizable parts, not reducible to $B\to P,V$ form factors.
In the following Sect.~2, I will discuss the form factors 
and in Sect.~3 the results of the recent analysis \cite{KMPW} of the charm-loop effect in $B\to K^{(*)} \ell^+\ell^-$ and $B\to K^*\gamma$.

\section{$B\to P,V$ form factors }
 The form factors of $\bar{s}_L\gamma_\mu b_L$ and $\bar{d}_L\gamma_\mu b_L$
currents involved in
$B\to P (V)\,\ell^+\ell^-$ decays are related, 
via $SU(3)_{fl}$ and isospin symmetry, respectively, to the form factors 
of $\bar{u}_L\gamma_\mu b_L$ current. The latter can  
in principle be determined from the measured $B\to\pi(\rho) \ell \nu_{\ell}$  semileptonic widths, with  
$|V_{ub}|$ taken from the inclusive semileptonic measurements.
However, for the $B\to K ^{(*)}$ form factors
this way of determination cannot be sufficiently accurate, 
because $SU(3)_{fl}$ is violated up to $\sim 20\%$
(like in the ratios $f_K/f_\pi$, $f_{DK}(0)/f_{D\pi}(0)$). 
The heavy-quark limit ($m_b, m_c\to \infty$) provides 
another useful flavour symmetry, predicting nontrivial relations between $B-$ and $D-$ meson hadronic amplitudes. Hence, in principle, one can 
try to obtain   
the $B\to P,V $ form factors employing the $D\to P,V$ form factors
extracted from the exclusive semileptonic $D$ decays. 
Again, to achieve a reasonable accuracy, one has to assess
the symmetry violating, $\sim 1/m_{c,b}$ corrections. They are 
generally not small. E.g., QCD calculations of the $B$ and $D$ decay 
constants yield $f_B\sim f_D\sim 200 ~\mbox{MeV}$, whereas in the heavy-quark limit 
 $f_D/f_B \sim \sqrt{m_b/m_c}\sim \sqrt{3}$. We come to a conclusion that in order
to reach $<20\%$ accuracy one needs 
a direct calculation of the $B\to P,V$  form factors in full QCD.

Currently, lattice QCD
with 3 dynamical flavours  has achieved a $\sim 10\%$  accuracy,  
but only for the $B\to \pi$ form factors 
in the region of  large $q^2>15 $ GeV$^2$. 
The future goal is to reach  5\% accuracy (see e.g.,\cite{latticetalk}). The 
$B\to K$ form factors have been obtained  recently 
\cite{QCDSF} in the quenched  approximation. 
The lattice calculations 
of $B\to V$ form factors ($V=\rho,\omega,K^*$) are complicated due to instability of vector mesons, and only earlier results
in quenched approximation are  available. 

In the important region of small and intermediate $q^2$
(large and moderate recoil of the final meson) the $B\to P,V$ form factors
are calculated from QCD light-cone sum rules (LCSR). 
This technique is used for finite quark masses and, therefore
takes into account the flavour-symmetry violation. The key 
nonperturbative objects are the light-cone
distribution amplitudes (DAs) of the light $P$- or $V$-meson.   
The LCSR method and results for $B\to \pi$ form factors 
are overviewed in \cite{talk_Ball}.
The most recent calculation for the $B\to \pi$ form factor is 
in \cite{DKMMO},  the $B\to K$ form factors were updated in
\cite{DM}. A typical uncertainty is about 15\%, with a little
room for improvement. The same method and input successfully reproduce the 
$D\to \pi$ and $D\to K$ form factors, as shown in \cite{KKMO}. 
The LCSR results for $B_{(s)}\to V$ 
form factors ($V=\rho$, $\omega$, $K^*,\phi$) are available from \cite{BZvect}.  
Note that the instability of
the vector meson  $V$ ( the $\rho\to \pi\pi$ or $K^*\to K \pi$ widths) 
are neglected also in the LCSR calculation.

Alternative LCSR's for $B\to P,V$ form factors are obtained 
with $B$-meson distribution amplitudes \cite{KMO}
taken as a nonperturbative input. Here all pseudoscalar and vector
mesons are treated on equal footing, being interpolated by a 
corresponding light-quark current. The  overall accuracy
of these sum rules is somewhat less than of 
the conventional LCSR's with DA's of light mesons. 
The gluon radiative corrections are not yet calculated 
and the uncertainties  of the parameters of $B$-meson DA's are still large.

Among the non-lattice tools for $B \to P,V$ form factors are also effective theories  (HQET, QCD factorization, SCET)
where non-trivial relations between the form factors
in the large recoil limit are predicted. A comprehensive analysis of $B\to K^*\ell^+\ell^-$ in terms of this approach can be found, e.g., in \cite{BFS}, where
the contributions with hard gluons to the 
decay amplitudes are identified and calculated.
The soft  $B\to P,V$ form factors  defined in the heavy-quark and large-recoil limit 
and the $B,P,V$-meson DA's  represent the external input. 
LCSR in SCET  \cite{LCSRscet} can be used to calculate the 
soft form factors.  Further increasing the accuracy in the effective theories
demands  taking into account the power-suppressed 
contributions.

In addition to the calculational methods, 
the analytical properties of the $B\to P,V$ form factors are employed,
in the form of ``series-parametrization''. 
The idea is to map the complex $q^2$-plane onto  the plane of the new variable $z(q^2)$, so that $|z|\ll 1$ in  semileptonic region $0<q^2<(m_B-m_{P(V)})^2$. Hence, a Taylor expansion  
around $z=0$ describes the form factor with a reasonable accuracy, allowing one
to inter/extrapolate the calculated form factor beyond the region of validity of lattice QCD or LCSR.  
The latest version of this parameterization was introduced in \cite{BCL} where  one can find all details. A recent analysis of all form factors  relevant for $B\to K^{(*)}\ell^+\ell^-$, 
combining LCSR and available lattice QCD results with series parameterization
can be found in \cite{BFW}. 

Summarizing, the current
uncertainty of $B\to P,V$ form factors is $ 12-15 \%$,
whereas $B\to V$ form factors have an 
additional ``systematic error'' related
to the instability of vector mesons.

\section{Charm loops in $B\to K^{(*)}\ell^+\ell^-$ }
\begin{figure}[h]
\begin{flushright}  
\includegraphics[height=.36\textheight]{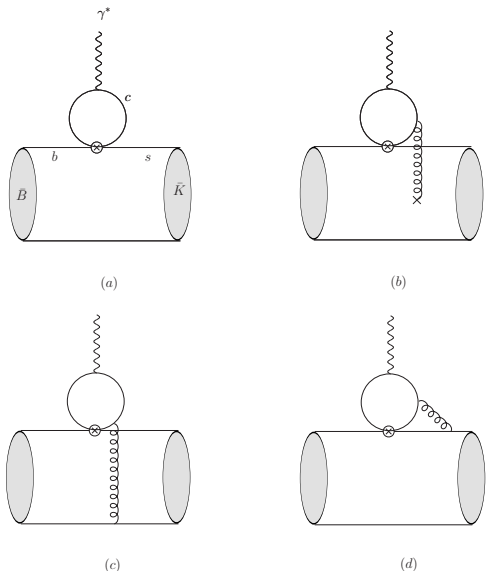}
\end{flushright}
\vspace{-3.4cm}
\caption{\it 
Charm-loop effect in $B\to K ^{(*)}\ell^+\ell^-$:\newline
(a)-the leading-order factorizable contribution;\newline (b)
nonfactorizale soft-gluon emission,\newline 
 (c),(d)-hard gluon exchange.
}
\end{figure}

In addition to the FCNC contributions containing $B\to P,V$ form factors,
the $B\to V(P) \ell^+\ell^-$ and $B\to V\gamma$ decay amplitudes 
are ``contaminated'' by the effects of weak interaction combined with e.m. interaction.
Let us discuss the most important ``charm-loop'' effect 
in $B \to K^{(*)} \ell^+\ell^-$ and $B \to K^* \gamma$, 
generated by the current-current operators 
$O_{1,2}^{(c)}$, acting
together with the $c$-quark electromagnetic current (see Fig.~1). 
In $B\to K^{(*)} \ell^+\ell^-$, this mechanism involves
an intermediate ``charm-loop", coupled to the lepton pair via the
virtual photon. In $B\to K^*\gamma$, the charm-loop is also possible 
if there is an additional gluon exchange with the rest of quarks. 

The simple $c$-quark loop diagram (Fig.~1a) is  usually
included in the factorization formula for $B\to
K^{(*)}\ell^+\ell^-$. In addition, hard-gluon exchanges between
the $c$-quark loop and the rest of the diagram (Fig.~1c,d) are taken into
account, together with other perturbative nonfactorizable effects
(see e.g., \cite{BFS}). One generally predicts these effects to be
small, if $q^2$ is far below the charmonium region.
The natural question is: how important are the contributions 
of the soft gluons emitted from the $c$-quark loop ? (Fig.1b)
A related question concerns the validity of the approximation
``$c$-quark-loop plus corrections'' at large $q^2$,
approaching the charmonium resonance region.
Note that at $q^2=m_\psi^2$, where $\psi=J/\psi, \psi(2S),...$
is one of the vector charmonium states,
the process $B\to K^{(*)}\ell^+\ell^-$ transforms into a
nonleptonic weak decay $B\to\psi K^{(*)}$, followed by the
leptonic annihilation of $\psi$. To avoid this ``direct'' charmonium background, the
$q^2$-intervals around $J/\psi$ and $\psi(2S)$ are subtracted from
the measured lepton-pair mass distributions in $B\to
K^{(*)}\ell^+\ell^-$. Nevertheless, the intermediate and/or virtual $\bar{c}c$
states contribute outside the resonance region and their effect
has to be accurately estimated.

In \cite{KMPW} these two questions were addressed, employing the 
expansion  near the light-cone of the product of the two operators:
$O^{(c)}_{1,(2)}$ and $c$-quark e.m. current.
As demonstrated in detail in \cite{KMPW}, this operator-product expansion
is valid at $q^2\ll 4m_c^2$, provided $2m_c\gg \Lambda_{QCD}$. 
The leading-order term of this expansion is reduced to the simple
$\bar{c}c$-loop, resulting in  the well-known loop function 
$g(m_c^2,q^2)$ multiplying the local operator
$\bar{s}_L\gamma^\rho b_L$.
The nontrivial effect is related to the one-gluon  term
which yields \cite{KMPW} a convolution 
\begin{eqnarray}
\widetilde{{\cal O}}_\mu(q) 
=  \int d\omega\, I_{\mu \rho \alpha\beta}(q,m_c,\omega) 
\bar{s}_L\gamma^\rho 
\delta [ \omega - {(i n_+ {\cal D}) \over 2}]
\widetilde{G}_{\alpha \beta} b_L 
\label{eq:oper}
\end{eqnarray}
of a nonlocal quark-antiquark-gluon operator  
with the calculable coefficient function $I_{\mu \rho \alpha\beta}$. 
In the above, $n_+ {\cal D}$ is the light-cone 
projection (defined so that $q\sim \frac{m_b}{2} n_+$
in the rest frame of $B$) 
of the covariant derivative acting on the gluon field-strength 
tensor and $\widetilde{G}_{\alpha \beta}=\frac{1}{2} \epsilon_{\alpha\beta\sigma\tau}G^{\sigma\tau}$.  
The explicit expression for this coefficient function 
is presented in \cite{KMPW}. As explained there
in more detail, two and more soft-gluon contributions are suppressed by 
additional powers of $1/(4m_c^2-q^2)$ with  respect  to
the leading one-gluon term. The operator in (\ref{eq:oper})
results from an effective resummation of the tower of local operators.
In the local limit, at $q^2=0$, we recover the local operator
of the charm-loop with soft gluon, taken into account 
first in \cite{Vol} for 
the $B\to X_s\gamma$ inclusive width and in \cite{KRSW}
for the $B\to K^* \gamma$ amplitude. 
\begin{figure}[h]
  \includegraphics[height=.2\textheight]{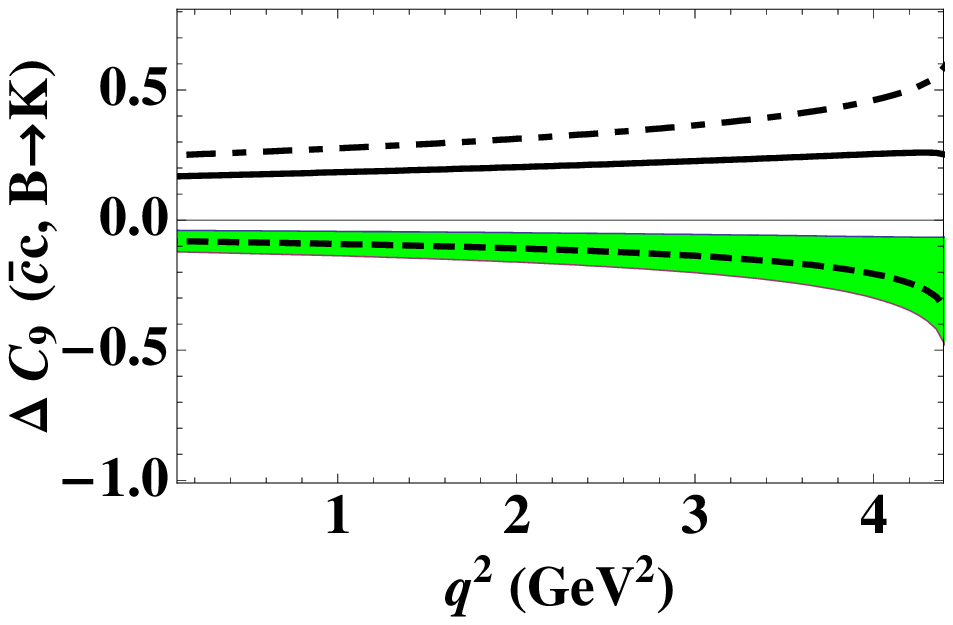}
\hspace{0.8cm}\includegraphics[height=.2\textheight]{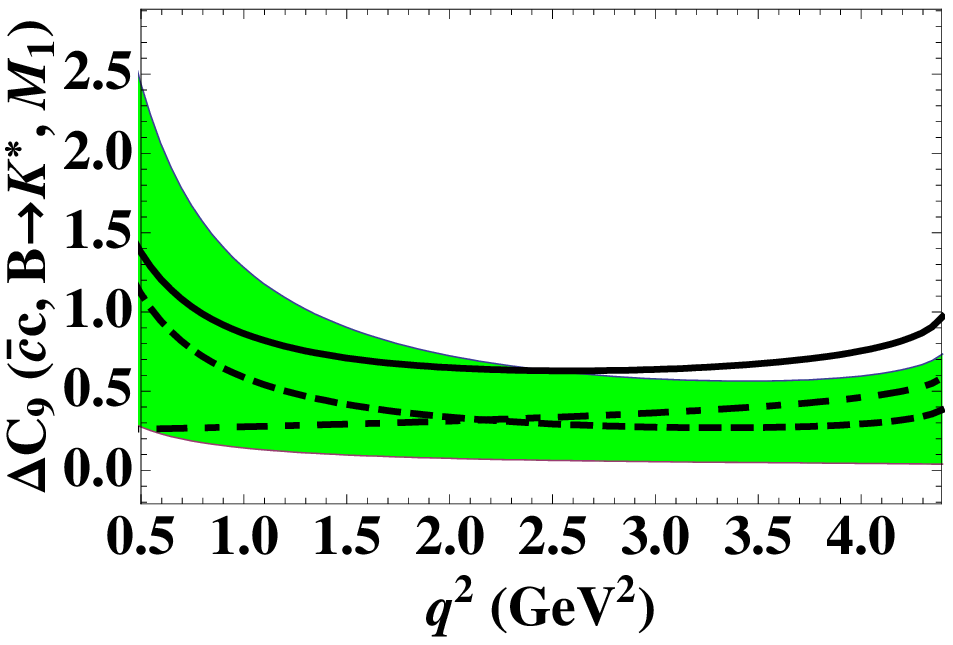} 
\caption{\it The charm-loop  effect in $B\to K \ell^+\ell^-$(left panel)
and $B\to K^* \ell^+\ell^-$(right panel, one of the three amplitudes)
expressed as a correction to the Wilson coefficient $C_9$ (solid),
including  the nonfactorizable soft-gluon contribution (dashed)
with the shaded region indicating the estimated uncertainty and
the factorizable contribution (dash-dotted).}
\end{figure}
In the adopted approximation, 
the calculation of the charm-loop effect at small $q^2$ 
is then reduced to the two hadronic matrix elements.
One of them is factorizable and expressed via $B\to K^{(*)}$ form factors.
The soft-gluon emission contribution yields a hadronic matrix element
of the nonlocal operator (\ref{eq:oper}).
This matrix element is calculated in \cite{KMPW} using the LCSR 
method \cite{KMO} where 
the $B$-meson DA's (approximated in HQET) are used 
as a universal nonperturbative input.

The result for the charm-loop contribution 
to $A(B\to K \ell^+\ell^-)$ including the soft-gluon
part is expressed in the form of a 
(process- and $q^2$-dependent) correction to the known 
Wilson coefficient $C_9$.
\begin{figure}[h]
\begin{flushright}  
\includegraphics[height=.22\textheight]{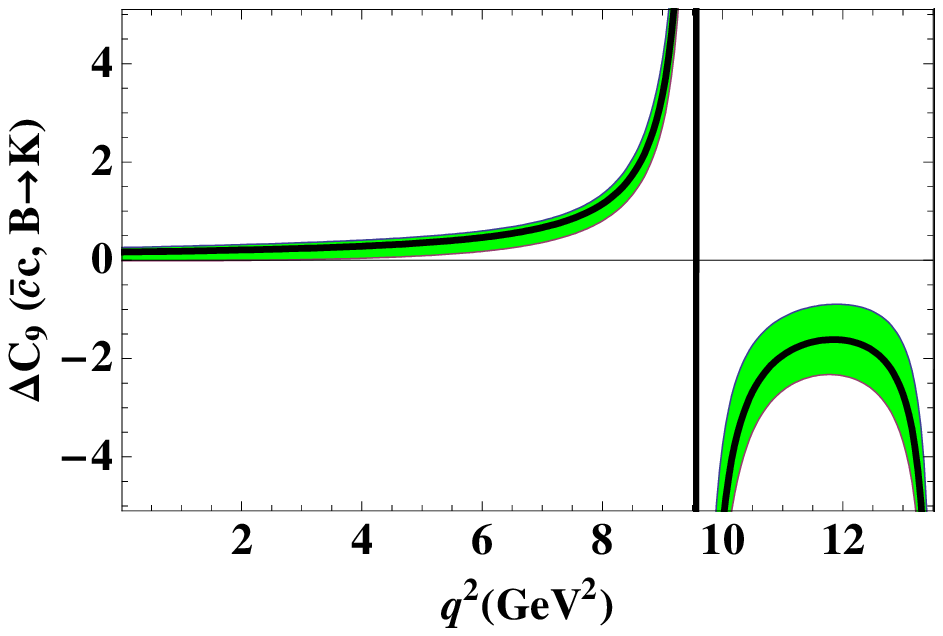}
\end{flushright}
\vspace{-4cm}
\caption{\it The charm loop  contribution\newline to the Wilson
coefficient $C_9$ for $\bar{B}_0 \to \bar{K} l^{+} l^{-}$ \newline at $q^2$
below the open charm threshold,\newline obtained from the dispersion
relation \newline fitted to the OPE result at $q^2\ll 4m_c^2$. 
\newline The central
values are denoted by dashed line, \newline shaded area indicates the
estimated uncertainties.}
\end{figure}
\begin{figure}[h]
\vspace{2cm}\includegraphics[height=.32\textheight]{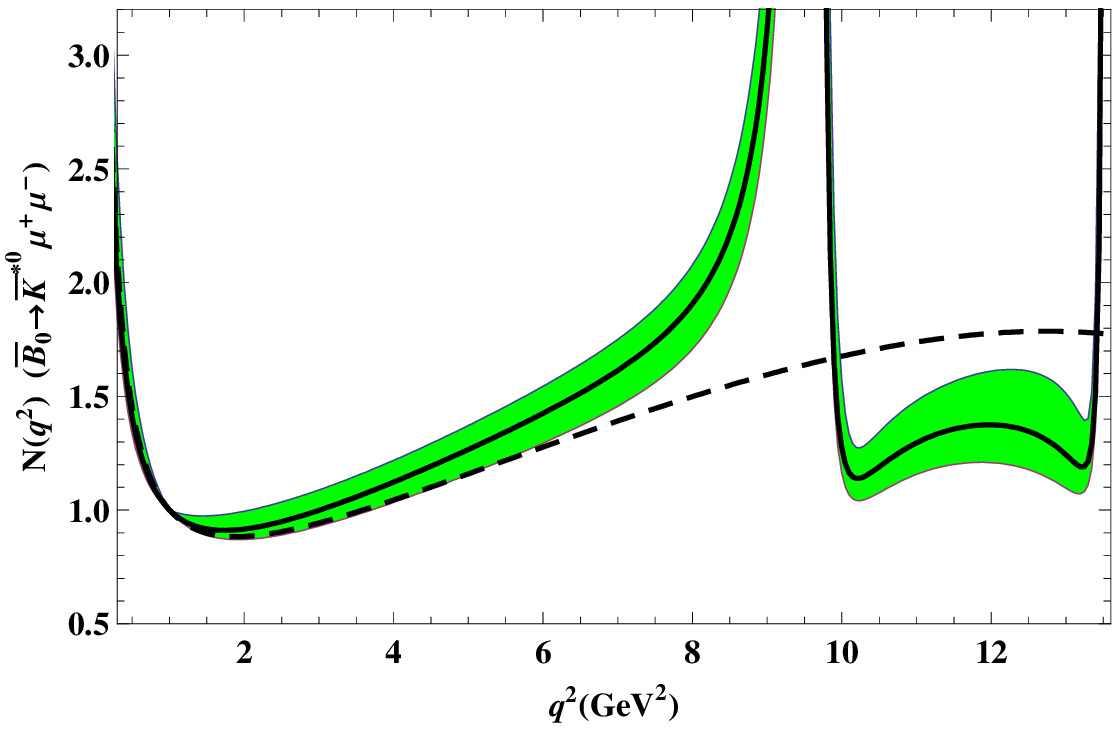} 
\hspace{1cm}
\includegraphics[height=.22\textheight]{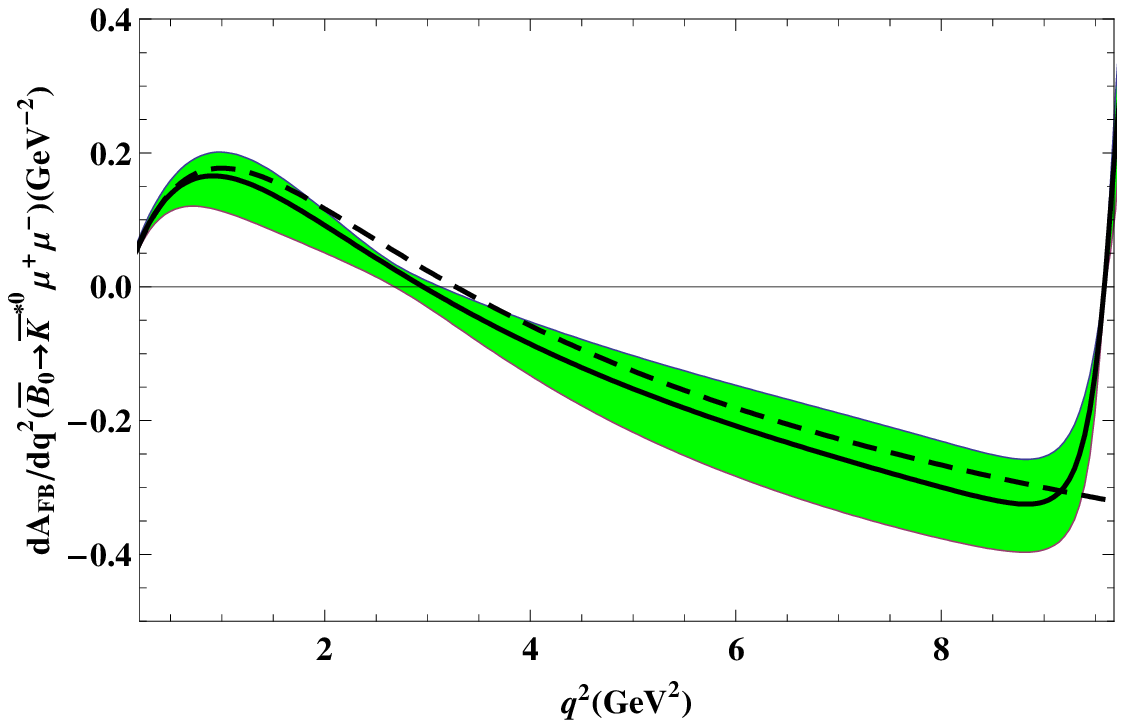}
\caption{\it left:The differential width of $\bar{B}_0 \to \bar{K}^\ast
\mu^{+} \mu^{-}$ normalized at $q^2=1.0$ {\rm GeV}$^2$,
including the charm-loop effect calculated with
the central values of input (solid, the shaded area indicates
estimated uncertainties) and without this effect (dashed); right:  The forward-backward asymmetry for $\bar{B}_0 \to
\bar{K^{\ast}} \mu^{+} \mu^{-}$ decay. }
\end{figure}

The calculated function $\Delta C_9^{(\bar{c}c,B\to K)}(q^2)$
plotted in Fig.~2[left] is valid at small $q^2\ll 4m_c^2$.
The numerical analysis reveals an important role 
of the soft-gluon part. It has an opposite sign
with respect to the factorizable loop term.   
The result 
\begin{equation}
\Delta C_9^{(\bar{c}c,B\to K)}(q^2=0,\mu\sim 
m_b)= 0.17^{+0.09}_{-0.18}\,,
\end{equation}
has to be added to  $C_9(\mu=m_b)$.

For $B\to K^* \ell^+\ell^-$ the effect is more 
pronounced and kinematically enhanced at small $q^2$
(see Fig.~2[right]). 
As a by-product of our calculation we also estimate 
the charm-loop effect in $B\to K^*\gamma$,
where the factorizable loop vanishes and only the  
nonfactorizable gluon emission contributes.

Furthermore, to access large $q^2$ we use the dispersion relation
in this variable for the invariant amplitudes
determining the $B\to K^{(*)}$ hadronic matrix elements,
saturating this relation with the first two charmonium levels.
This relation is valid at any $q^2$, hence we can match 
it to the result of QCD calculation  
at $q^2\ll4m_c^2$. In addition, we fix
the absolute values of the residues 
from experimental data on $B\to \psi K$ widths. The integral 
over the spectral density of higher states is 
then fitted as an 
effective pole. After fixing the parameters of the dispersion relation
we predict the correction $\Delta C_9(q^2)$ at large $q^2$
(see Fig.~3).

Finally, the observables in $B\to K^{(*)}\ell^+\ell^-$ are calculated 
employing the form factors and charm-loop amplitudes
(see Fig.4). There is a moderate influence of the charm-loop effect 
on the position of the zero in the forward-backward asymmetry.

Concluding, I would like to
emphasize that a careful analysis of all other 
similar effects (light-quark loops, weak annihilation etc.) including soft-gluon contributions is necessary for obtaining a complete and accurate prediction for 
$B\to V(P)\ell^+\ell^-$ and $B\to V\gamma$ in SM.
\vspace{-0.3cm}
\Acknowledgements
I am grateful to Martin Gorbahn and Yu-Ming Wang for useful comments.
This work is supported by the Deutsche Forschungsgemeinschaft
under the  contract No. KH205/1-2.


\begin{thebibliography}{99}

\bibitem{BBL}
  G.~Buchalla, A.~J.~Buras and M.~E.~Lautenbacher,
  Rev.\ Mod.\ Phys.\  {\bf 68} (1996) 1125.

\bibitem{BFS}
  M.~Beneke, T.~Feldmann and D.~Seidel,
  Nucl.\ Phys.\  B {\bf 612} (2001) 25.

\bibitem{KMPW}
  A.~Khodjamirian, T.~Mannel, A.~A.~Pivovarov and Y.~M.~Wang,
  JHEP {\bf 1009} (2010) 089.


\bibitem{latticetalk} J.~Shigemitsu, in these proceedings

\bibitem{QCDSF}
  A.~Al-Haydari {\it et al.}  [QCDSF Collaboration],
  Eur.\ Phys.\ J.\  A {\bf 43} (2010) 107.

\bibitem{talk_Ball} P.~Ball, in these proceedings

\bibitem{DKMMO}
G.~Duplancic, A.~Khodjamirian, T.~Mannel, B.~Melic and N.~Offen,
JHEP {\bf 0804}, 014 (2008).

\bibitem{DM}
  G.~Duplancic and B.~Melic,
  Phys.\ Rev.\  D {\bf 78}, 054015 (2008).


\bibitem{KKMO}
A.~Khodjamirian, C.~Klein, T.~Mannel and N.~Offen,
Phys.\ Rev.\  D {\bf 80}, 114005 (2009).


\bibitem{BZvect}
P.~Ball and R.~Zwicky,
Phys.\ Rev.\  D {\bf 71}, 014029 (2005).


\bibitem{KMO}
  A.~Khodjamirian, T.~Mannel and N.~Offen,
  Phys.\ Rev.\  D {\bf 75} (2007) 054013.


\bibitem{LCSRscet}
  F.~De Fazio, T.~Feldmann and T.~Hurth,
  JHEP {\bf 0802}, 031 (2008).


\bibitem{BCL}
C.~Bourrely, I.~Caprini and L.~Lellouch,
Phys.\ Rev.\  D {\bf 79} (2009) 013008.


\bibitem{BFW}
  A.~Bharucha, T.~Feldmann and M.~Wick,
  JHEP {\bf 1009} (2010) 090.




\bibitem{Vol}
M.B.~Voloshin,
Phys.\ Lett.\ {\bf B397} (1997) 275.

\bibitem{KRSW}
A.~Khodjamirian, R.~Ruckl, G.~Stoll and D.~Wyler,
Phys.\ Lett.\ {\bf B402} (1997) 167.

\end{thebibliography}
\end{document}